\documentclass[a4paper,conference,twocolumn,10pt]{IEEEtran}

  \usepackage[pdftex]{graphicx}
\usepackage{placeins}
\usepackage[nolist,nohyperlinks]{acronym}
\usepackage{color}
\usepackage{amsmath}
\usepackage{amssymb}
\usepackage{amsthm}
\usepackage[pdftex]{hyperref}
\usepackage{cleveref}
\usepackage{booktabs}
\usepackage[sort&compress]{natbib}
\usepackage{multirow,array,varwidth}
\usepackage{subfig}
\usepackage{verbatim}
\usepackage{MnSymbol}
\usepackage{listings,xcolor}
\usepackage{balance}
\usepackage{float}
\usepackage[singlespacing]{setspace}

\bibpunct{[}{]}{,}{n}{,}{;}
\definecolor{myRed}{rgb}{1,0,0}
\definecolor{myGreen}{rgb}{0,1,0}
\definecolor{myBlue}{rgb}{0,0,1}
\crefformat{chapter}{#2Chapter~#1#3}
\crefformat{section}{#2Section~#1#3}
\crefformat{subsection}{#2Section~#1#3}
\crefformat{Appendix}{#2Appendix~#1#3}
\crefformat{figure}{#2Figure~#1#3}
\crefformat{table}{#2Table~#1#3}
\crefformat{equation}{#2(#1)#3}
\crefmultiformat{equation}{#2(#1)#3}{ and~#2(#1)#3}{, ~#2(#1)#3}{, and~#2(#1)#3}
\crefmultiformat{figure}{#2Figures~#1#3}{ and~#2#1#3}{, ~#2#1#3}{, and~#2#1#3}
\crefmultiformat{table}{#2Tables~(#1)#3}{ and~#2(#1)#3}{, ~#2(#1)#3}{, and~#2(#1)#3}
\crefrangeformat{equation}{#3(#1)#4--#5(#2)#6}
\crefrangeformat{figure}{#3Figures~#1#4--#5#2#6}
\crefrangeformat{section}{#3Sections~#1#4--#5#2#6}
\crefname{lemma}{Lemma}{Lemmas}

\begin{document}

\begin{acronym}[DSTTD-SGRC]
\acro{CDF}{Cumulative Distribution Function}
\acro{GIG}{Generalized Integer Gamma}
\acro{LTE}{Long Term Evolution}
\acro{LTE-A}{LTE-Advanced}
\acro{MRC}{Maximum Ratio Combining}
\acro{MRT}{Maximum Ratio Transmission}
\acro{RU}{Remote Unit}
\acro{RV}{Random Variable}
\acro{SIR}{Signal-to-Interference Ratio}
\acro{UE}{User Equipment}
\acro{CBP}{Codebook-Based Precoding}
\acro{HST}{High Speed Train}
\acro{BS}{Base Station}
\acro{RoF}{Radio over Fiber}
\acro{LOS}{Line of Sight}
\acro{GPS}{Global Positioning System}
\acro{ISM}{Industrial, Scientific and Medical}
\acro{EIRP}{Equivalent Isotropically Radiated Power}
\acro{CoMP}{Coordinated Multipoint}
\acro{CAPEX}{Capital Expenditure}
\acro{OPEX}{Operational Expenditure}
\acro{ICI}{Inter Carrier Interference}
\acro{OFDM}{Orthogonal Frequency Division Multiplexing}
\end{acronym} 

\title{Providing Current and Future Cellular Services to High Speed Trains}
\author{
    \IEEEauthorblockN{Martin Klaus M\"uller, Martin Taranetz and Markus Rupp\\[2mm]}
    \IEEEauthorblockA{Institute of Telecommunications, TU Wien\\
                  Gusshausstrasse 25/389, A-1040 Vienna, Austria\\
                  Email: \{mmueller, mtaranet, mrupp\}@nt.tuwien.ac.at\\
                  }
}

\maketitle


\newpage

\begin{abstract}
The demand for a broadband wireless connection is nowadays no longer limited to stationary situations, but also required while traveling. Therefore, there exist combined efforts to provide wireless access also on High Speed Trains (HSTs), in order to add to the attractiveness of this means for transportation. Installing an additional relay on the train, to facilitate the communication, is an approach that has already been extensively discussed in literature. The possibility of a direct communication between the base station and the passenger has been neglected until now, despite it having numerous advantages. Therefore, a comparison between these two opposing approaches is presented in this paper, accompanied by a detailed discussion of the related aspects. The focus is set on the feasibility of the direct link approach, including simulation results. Further technical issues are also presented, especially regarding the interdependencies of the different aspects and providing a view of mobile- and train-operators on the topic.
\end{abstract}


\acresetall

\section{Introduction}
\label{Sec:Introduction}
In the current market of train services, being able to provide mobile broadband access to costumers has become a main inducement for choosing this means of transportation. Due to the ubiquitous use of the Internet and the rapid adoption of novel devices such as smart phones and tablet computers, most passengers have become accustomed to experiencing high data rates and having the service following them no matter where they go. With the number of commuters being expected to increased, the high user mobility is also one of the most emphasized scenarios in the initiative for the fifth generation of wireless communications (5G). \ac{LTE-A}, the contemporary standard for wireless communication, is not optimized for the challenges of \ac{HST} scenarios. Hence, many train operators - mostly in collaboration with mobile operators - have increased efforts to satisfy the ever increasing requirements. There also exist international collaborations on a broad level to push for higher data rates and shorter latencies, for example the Shift2Rail initiative by the European Union \cite{shift2rail}.

Wireless communications in \ac{HST} scenarios is confronted with unique conditions, which have a considerable impact on network planning. In particular, the scenarios are characterized by \acp{UE} being densely packed inside of the train and moving at high speed, as well as the specific propagation effects in a diversity of different environments.

Most publications on this topic assume a relay-based approach, assuming additional hardware installed on the train, that communicates with the \ac{BS} as well as with \acp{UE} without them communicating directly (\cite{5872303}, \cite{penetration_values}). However, the direct communication between \acp{UE} and \acp{BS} has its own advantages, but has not been given enough attention in literature. This contribution provides an extensive comparison between the relay approach and the less studied direct link approach and also discusses various other aspects that are peculiar for \ac{HST} scenarios.

The structure of the paper is the following: \cref{Sec:Train_Issues} explains general issues that arise in \ac{HST} scenarios. \cref{Sec:Relay_vs_Direct} contains the comparison of relay and direct-link approaches. \cref{Sec:Further_Aspects} then deals with further aspects of \ac{HST}, showing dependencies between different parameters and discussing the system aspects from an operator's point of view. Conclusions are drawn in the final section.
\section{Special train issues}
\label{Sec:Train_Issues}

\begin{figure*}
	\centering
	\includegraphics[width=1.8\columnwidth]{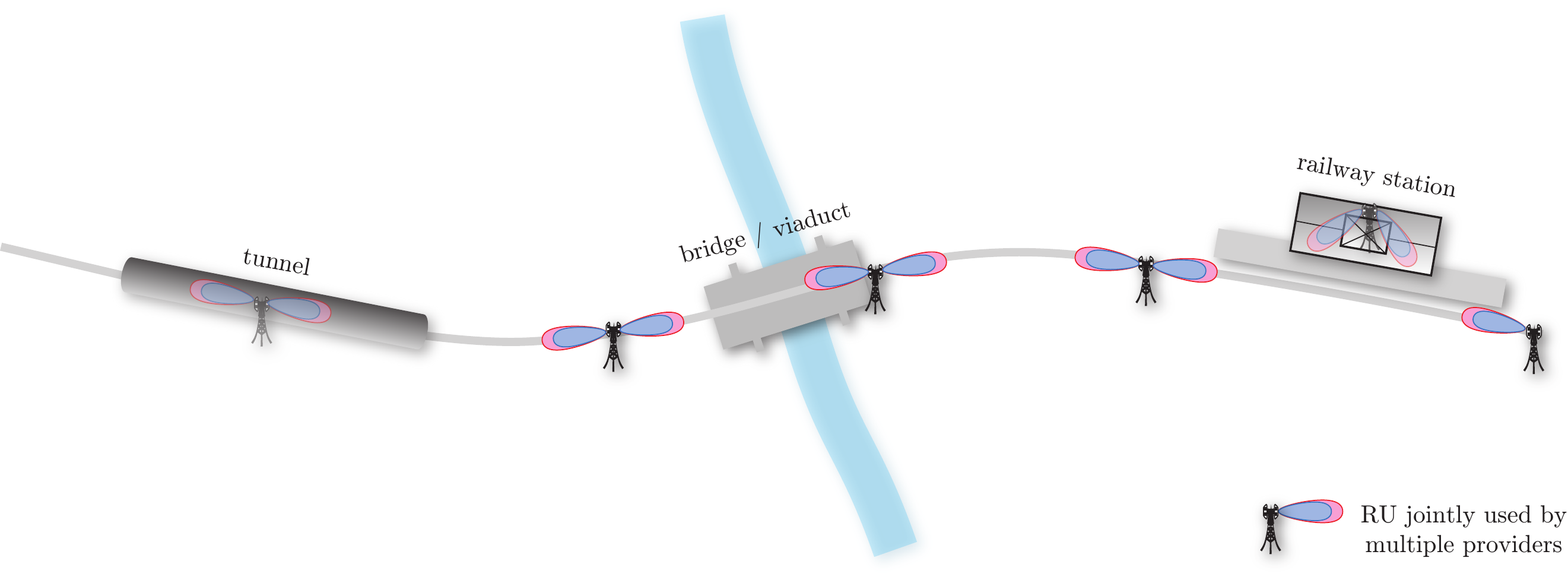}
	\caption{Various environments along railroad track. The track is supplied by \acp{RU} that are used by multiple network operators.}
	\label{Fig:RU_setup}
\end{figure*}

Wireless communication in \ac{HST} scenarios exhibits several key differences to the traditional considerations of a coverage oriented network. What is more, there is not the same amount of extensive experience with such scenarios as with classical networks. Thus, many effects have not been fully understood and it is not clear which are the most significant in \ac{HST} scenarios. The state of the art is discussed in the following, along with the aspects that are currently identified as most important.\\

An intrinsic characteristic is the high speed of the train that can exceed values of 350\,km/h. This leads to a high Doppler shift, which causes several transceiver impairments such as channel estimation errors and \ac{ICI} in \ac{OFDM} systems.

Moreover, the channel characteristics along the tracks vary greatly. The surroundings show several distinct categories, which encompass certain different features that influence the signal propagation. Among those are tunnels, trenches, cuttings, stations, viaduct-like structures or bridges. An example for such a diverse environment is shown in \cref{Fig:RU_setup}. An extensive description of such categories is described in \cite{aichallenges}. In addition to the onerous propagation conditions, a large part of the signal may be shielded by the metal chassis of the carriage. If the signal is first received by a relay, this can lead to a very strong multi-path component additionally to the \ac{LOS} path.

Another peculiarity of \ac{HST} scenarios is the \ac{UE} distribution and -movement. Opposed to classical assumptions of random \ac{UE} placement, all \acp{UE} are concentrated within the train. Their location and movement is approximately deterministic as it is predefined by the course of the tracks and the speed of the train. This implicates a certain a-priori knowledge that can be exploited to compensate for some of the above mentioned issues.

The high \ac{UE} concentration results in pulse-like traffic in the cells along the tracks. This leads to \acp{BS} being either completely idle or confronted with a high load (in case the hardware is dedicated to serving the train) or to large performance fluctuations in a cell (in case also the surrounding environment is covered by the \acp{BS}). From the high speed of the train it follows that small cells are traversed in a very short time, thus leading to a frequent necessity for handovers. When all \acp{UE} aboard a train simultaneously require a handover to the next cell, the control channel easily gets congested and suspends \acp{UE} from associating with the neighboring cell. As indicated above, the deterministic \ac{UE} behavior may provide a considerable advantage for implementing elaborated handover schemes.

It is commonly agreed that a satisfactory quality of service can only be achieved by employing dedicated hardware in an \ac{HST} scenario, especially in the form of \acp{RU}. Due to their smaller size and cost compared to classical \acp{BS}, it is possible to install them close to the tracks and in great numbers. Several \acp{RU} can be connected to one \ac{BS} via \ac{RoF}. This allows for very flexible routing options as explained in the subsequent section. An example for an \ac{RU} installation along the tracks is presented in \cref{Fig:RU_setup}. In the remainder of this work, we will assume such scenario for our considerations.

Due to the diversity of environments, the modeling of the fast fading channel becomes a complex task. Many considerations regarding wireless communication in \ac{HST} scenarios are based on simulations applying the Winner channel model \cite{winner2model}. Even though the model is highly adjustable, it was originally not intended to represent the characteristics of an \ac{HST} scenario. Furthermore, it does not reflect the dynamic changes in the channel characteristics from one category to another. Thus, such simulations only yield first order statements that may considerably deviate from reality and may not appropriately reflect the specifics of the environment. 3GPP has recently introduced a 3D channel model in \cite{3gpp36873}. However, specific aspects for train communications are not yet included and have to be determined.

\section{Relay vs. Direct Link }
\label{Sec:Relay_vs_Direct}

\begin{figure*}
	\centering
	\includegraphics[width=1.8\columnwidth]{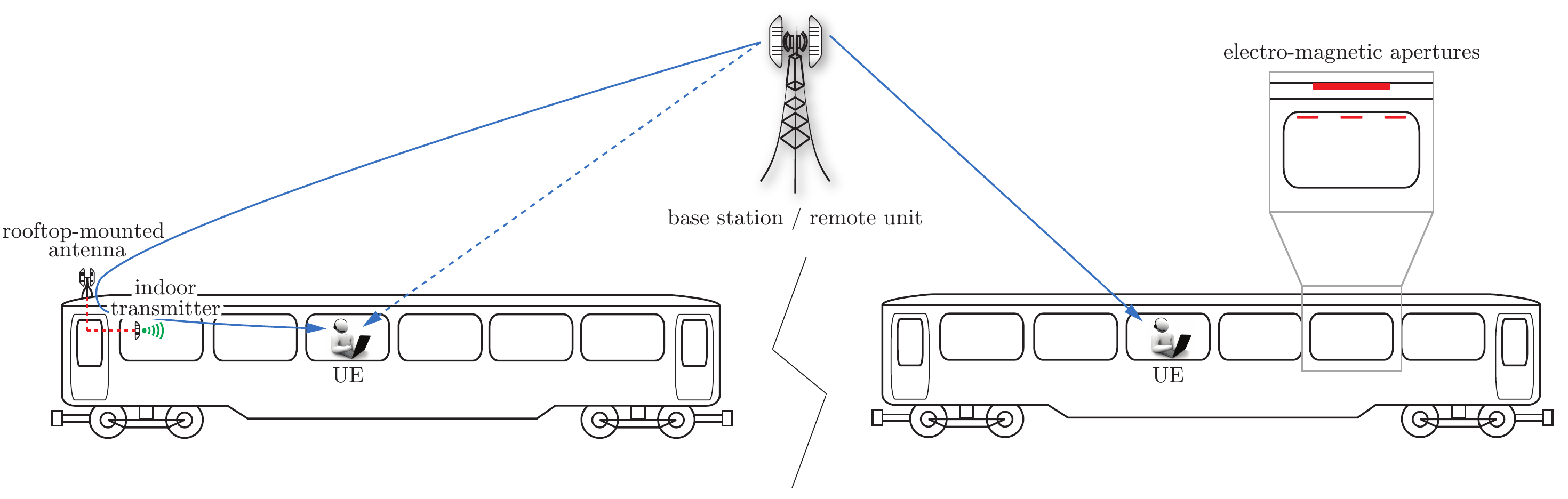}	
	\caption{Signal reception with relay- and direct-link approach. Zoomed view indicates possibilities for applying electro-magnetic apertures.}
	\label{Fig:Relay_Direct}
\end{figure*}

In this section we consider the downlink direction exclusively. Similar considerations are valid for the uplink direction.
Generally speaking, there exist two opposing approaches to provide wireless communications to passengers of an \ac{HST}. In the first case, the UE directly associates with the \acp{BS} along the tracks, while in the second case this link is established via a relay, as shown in \cref{Fig:Relay_Direct}. Subsequently, a comparison between these two approaches is drawn and advantages and drawbacks on both sides are discussed.

\subsection{Relay Approach}
In the relay scenario, one or several antennas are mounted on the outside of the train. These are connected to one or more relays which are then distributing the signal inside the train. This approach has the major advantage that the signal is not attenuated by the windows of the carriage. Moreover, this setup allows to configure the relay such that it appears as a single \ac{UE} to the \ac{BS}, thus significantly reducing the number of handovers. Therefore, all traffic is aggregated by the relays and then distributed to the \acp{UE}.

While the quality of the experienced link might considerably be improved, the employment of relays does not come without cost. Firstly, relays have to be licensed for the specific band they are operating on. This is of minor consequence when no borders are crossed. In smaller countries such as, e.g., in most European, one train connection can easily span three or more countries. For each country, the relays have to be registered individually or have to be switched off, which leaves the system in a direct-link state. Another consequence is, that the employment of carriages becomes restricted to the countries/routes for which the relays on board are licensed and thus becomes less flexible.

A second issues appears with the choice of the frequencies for the \ac{RU}-to-relay and the relay-to-\ac{UE} connection. If the same frequency is used on both link-sections, thus only bypassing the penetration loss of the carriage, the \ac{UE} might still receive a considerable amount of the desired signal by a direct link through the window. The relative receive power of these two links depends on the individual position of the \ac{UE} and the actual penetration loss. Due to the latency caused by the relay, the signal might be perceived via several multi-path components that cannot be equalized.

The aforementioned problem is completely avoided when two different frequencies are used on both sections of the link. For example, a dedicated frequency of a mobile operator is used from \ac{RU} to the relay and a second frequency, e.g., in the ISM band, for the supply with WiFi inside. This setup is frequently referred to  in literature \cite{5872303}. Nonetheless, it only provides a data connection for the passengers. Since only the relay is visible as a single \ac{UE} for the \ac{BS}, the passengers are not accessible for mobility management. 

The performance of this setup may also considerably depend on the number of antennas and relays per train/carriage. It has to be scrutinized whether the relays should work individually (e.g., one per carriage) or if the received signals should be combined. This is also affected by the possibility to connect all relays to all carriages. Since a cable connection will not be feasible in practice, near field communication standards may be considered for this task.

\subsection{Direct-Link Approach}

The direct-link approach assumes a direct connection between \ac{RU} and \ac{UE}. In comparison to the above scenario, the signal does experience a severe penetration loss into the carriage in this case. As the chassis of the carriages is usually made of metal, the signal enters the train mainly through the windows. However, the penetration loss may greatly vary among window types, as they are mostly metal coated themselves.  Attenuation values range from 10\,dB to 40\,dB (see \cite{penetration_values}).

Note that these values reflect the situation for current carriages in use. Since the interest among train operators is  increasing to provide best quality of experience to their customers, the design of future trains is likely to be adapted to the demands of the wireless link. Among various options is the possibility to introduce windows with small penetration loss (omitting the metal coating). Another option is to include apertures in the chassis or the window itself, by incorporating materials that are more permeable for electro-magnetic waves, for example carbon fiber materials. Examples for such apertures are indicated in \cref{Fig:Relay_Direct}.

System simulations were performed in order to demonstrate that the direct-link setup is feasible with \ac{LTE-A} (as mentioned, it was not designed for this task) even under worst case conditions of high penetration loss. A scenario with four equidistantly spaced sites that are placed directly next to the train tracks (thus exploiting available infrastructure) is considered. Each site employs two \acp{RU} pointing in opposite directions along the tracks. \acp{RU} facing away from the train are omitted in the analysis. Three \ac{RU} collaboration schemes are compared: (i) each site is connected to an individual \ac{BS} (baseline scheme), (ii) two dominant \acp{RU} coordinate their transmission such that interference from one \ac{RU} is omitted (coordination scheme), and (iii) the two dominant \acp{RU} are connected to the same \ac{BS} (cooperation scheme). A typical German ICE-train fully occupied with 460 passengers and 10\% having an active wireless connection is regarded. A penetration loss of 30\,dB is applied referring to a worst case experience. Further simulation results can be found in \cref{Tab:Simulationparameters}. Simulations were performed with the Vienna \ac{LTE-A} Downlink System Level Simulator \cite{ltesim2014} (current version v1.8 r1375).

\begin{table}
	\centering        
	\caption{Simulation Parameters.}\label{Tab:Simulationparameters}  
	\begin{tabular}{r|l}
		\toprule
		\textbf{Parameter} & \textbf{Value} \\
		\hline
		System bandwidth & 20\,MHz \\
		Carrier frequency & 2.14\,GHz \\
		Inter-\ac{RU} distance & 1000\,m \\
		eNodeB transmit power $P_{\rm T}$ & 40\,W\\
		Antennas per \ac{RU} & 2 \\
		MIMO mode & CLSM \\
		Path loss model & TS 36.942 'rural'  \cite{3gppTS36942}\\
		Channel model &  ITU-R Vehicular A \cite{ItuRVehA} \\
		Train speed & 200\,km/h\\
		Receiver type & zero forcing \\
		Noise power spectral density & - 174\,dBm/Hz \\
		Receiver noise figure & 9\,dB \\
		Train length & 200.84\,m \\
		Active \acp{UE} & 46 \\
		\ac{UE} distribution within train & random, according to a\\
		& uniform distribution \\ 
		Antennas per \ac{UE} & 1 \\
		Traffic model & full buffer \\
		Scheduler & proportional fair \\
		Channel knowledge & perfect \\
		Feedback & AMC: CQI, \\
		&  MIMO: PMI and RI \\
		\ac{RU} backhaul connection & radio over fiber, no delay \\
		\bottomrule
	\end{tabular} 
\end{table}

\begin{figure}
	\centering
	\includegraphics[width=.95\columnwidth]{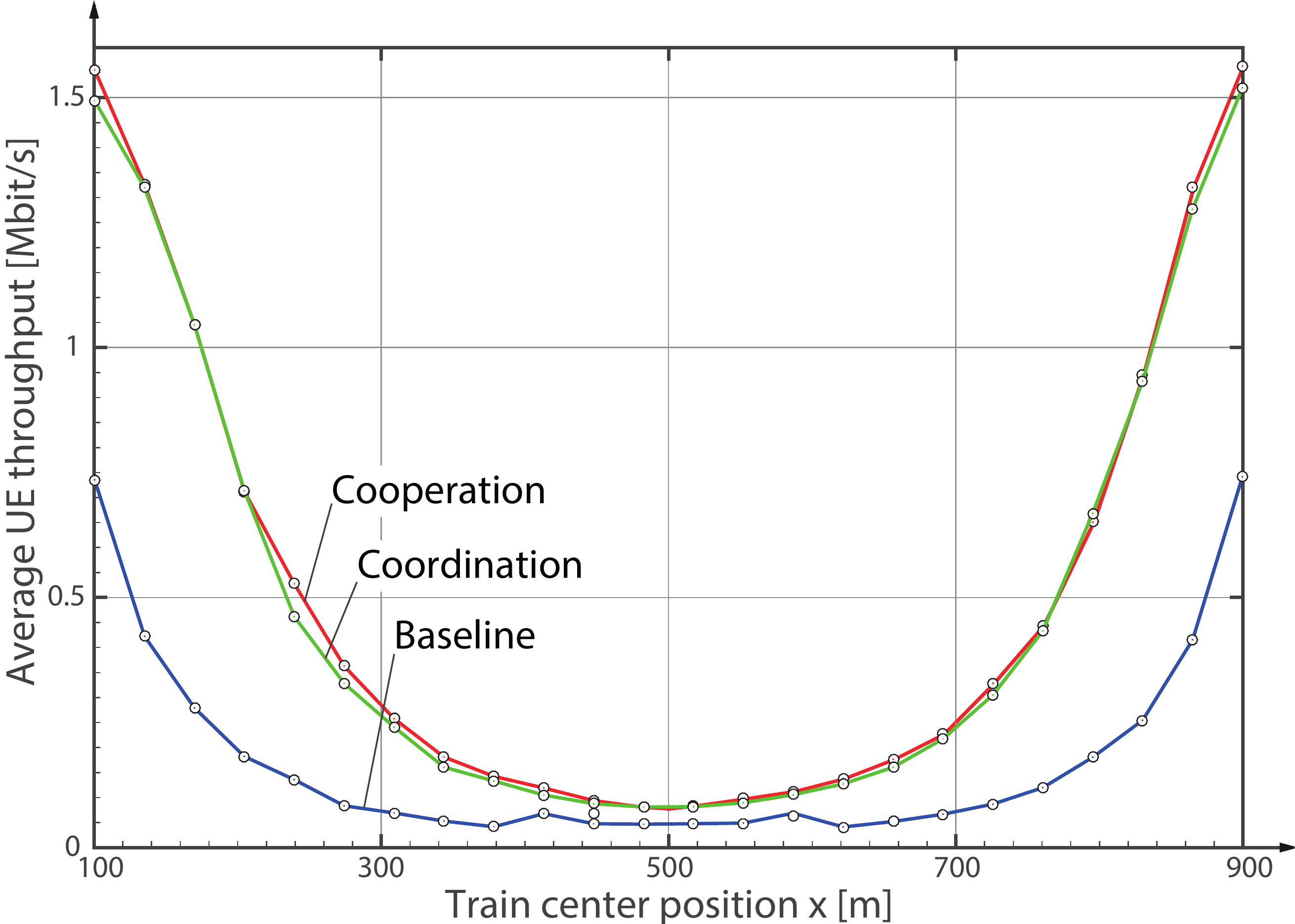}
	\caption{Train average throughput [Mbit/s] versus train center position. Curves refer to results for baseline-, coordination- and cooperation scheme among \acp{RU}. Circles indicate 95\,\% confidence intervals.}
	\label{Fig:Sim_Results}
\end{figure}

Simulation results are shown in \cref{Fig:Sim_Results}. It is observed that sophisticated collaboration schemes can improve the system performance. The U-like shape of the curves could be exploited in combination with a sophisticated scheduler. In regard of the traffic-type, \acp{UE} with delay sensitive data are assigned resources such that their requirements are fulfilled, while \acp{UE} with best-effort traffic models are mostly served when the train is closer to an \ac{RU} and a higher total data rate is available. The curve corresponding to the baseline scenario is not flat in the middle, since starting from 400\,m, fractions of \acp{UE} are already assigned to the next \ac{RU}, thus improving the average throughput (correspondingly at 600\,m).\\

While it is possible in the relay approach to let the whole train appear as the equivalent of one or two \acp{UE} (given the proper frequency setup), this option is not available for the direct-link approach. Each active passenger appears as an individual \ac{UE}. Therefore, many handovers have to be executed when the train moves from one cell to another. Considering the high speeds and that even "inactive" passengers, not actively transmitting data, have to be handed over to the next cell, this can lead to a considerable amount of traffic on the control channel and might in extreme cases lead to blocking of \acp{UE}.\\

As mentioned above, the deterministic location of the train and the semi-deterministic location of the UEs (confined inside of the train, but not known in particular) enables new concepts of handover and cell extension, including

\begin{itemize}
	\item Moving cell
	\item Smart handover schemes
\end{itemize}

The moving cell concept is a well known concept. For both approaches, relay and direct-link, it increases the cell length from around 1\,km to a maximum of up to 50\,km. Commercial systems with the necessary capabilities for that task are already available (e.g., \cite{kbow}). The working principle is that many \acp{RU} are associated with a central control entity via \ac{RoF}. When the train leaves the range of one \ac{RU}, the central unit will reroute the data-stream towards the next \ac{RU} along the tracks. In that manner, the cell 'travels' with the train (in an abstract sense, this can be interpreted as a semi-static beamforming scheme), enabling a transparent and smooth transition, without the need for handovers over longer distances. Considering a speed of 350\,km/h, the transition of a cell of 1\,km length takes approximately 10 seconds and a handover becomes necessary. With the moving cell concept, the period of reoccurring handovers is enhanced to 9 minutes. The concept is further detailed in \cite{917517}.\\

Smart handover schemes were reported, e.g., in \cite{handover}. Such schemes improve the handover performance, but assume cells simply touching each other on the outer cell borders.\\


\section{Further Aspects}
\label{Sec:Further_Aspects}

The aspects of wireless communications in \ac{HST} scenarios that have been discussed in the previous sections considerably affect many design choices for an actual system. The choice between relay- and direct link approach is aggravated by the fact that many of the presented aspects are interdependent, thus influencing each other. In this section, these dependencies are discussed from the train- and the mobile-operators' point of view.\\

\subsection{Technical Aspects}
The interior of a train is characterized by many objects such as seats and baggage racks as well as passengers that reflect or absorb electro-magnetic waves. For a direct-link as well as for the relay approach, this has to be taken into account. The actual position of the \ac{UE} might have an even greater impact than the characteristics of the outdoor link. In the relay approach, only the characteristics of the interior of the carriage have to be considered for the second part of the link. For the direct-link, a possible loss due to penetration through compartment walls need to be considered. Such path can even be dominant over one passing directly through the window closest to the \ac{UE}, since the penetration through windows might show a strong dependence on the angle of incidence. This has to be considered in the modeling of the direct-link channel. Possible shadowing of a train passing by in the opposite direction constitutes a further issue, that might be included as a possible environment scenario, which occurs with low probability.

The selection of the carrier frequency also considerably impacts the overall system-performance. The most relevant options are:
\begin{itemize}
	\item Licensed bands (i.e., \ac{LTE-A} frequencies)
	\item Unlicensed bands (\ac{ISM} bands)
	\item free bands in mmWaves
\end{itemize}
On the one hand, licensed bands allow the access to mobility management. On the other, distinct bands might experience a considerably different interference environment. In rural scenarios, the interference from neighboring \acp{BS} can become a major issue, while for \ac{ISM} bands, due to the limited \ac{EIRP}, and due to the lack of close interferers, the interference will be smaller. The limitation of the radiated power could be compensated with the deployment of very cost efficient \acp{RU} along the tracks, being spaced at very short distances. The application of mmWave is currently a hot topic for dense static deployment scenarios, however their applicability for high mobility scenarios has yet to be scrutinized. A possible application could be to radiate mmWaves from the tracks (i.e., from surface level) to the train base (assuming antennas being installed there).

The choice of a relay- or a direct-link approach also strongly impacts the optimal placement and orientation of the \acp{RU}. While transmitter directions being aligned along the tracks work well in combination with relays, a positioning off the tracks with an antenna orientation perpendicular to the track-orientation can be beneficial for the direct-link approach, as it may overcome issues with a shallow angle of incidence, causing high penetration loss through the windows.

\subsection{Operator View}
Besides all the technical aspects, mobile operators will also take economical factors such as \ac{CAPEX} and \ac{OPEX}, political issues, necessary cooperations and complexity of implementation into consideration. The two key players in an \ac{HST} scenario are the train operators and the mobile operators, whose perspectives might considerably deviate from each other.\\

For a mobile operator, the direct-link comes with the benefit of not having to rely on additional hardware being installed on the train. Thus, their system for supplying wireless access becomes more independent. The train-types only have to grant the expected penetration characteristics. Additionally, the mobile operator has all options for mobility management as opposed to a relay scenario with WiFi on the second part of the link. A train operator might also favor the direct approach, once for not being obligated to install and maintain additional hardware, but more importantly for not having to deal with any legal issues when trains are crossing borders. For both sides this approach comes with the benefit of a reduced necessity for synchronizing with each other.

Along with this first decision comes the question of average performance, that should be supplied to the passengers. Most importantly this affects the optimal placement and distance of the \acp{RU}. The total amount of \acp{RU} and connected hardware along the tracks (e.g. control entities, acquired sites) defines the \ac{CAPEX} and \ac{OPEX} of the whole system. Thus, the \ac{RU} spacing needs to be optimized for granting the required performance while minimizing the cost. Modifying the carriages might therefore change the cost picture completely, but thereby offloading some of the cost to the train operator. For example, installing windows with a low penetration loss or carriages with electro-magnetic apertures and employing the direct-link approach will enable to significantly increase the spacing of the \acp{RU}, thus reducing \ac{CAPEX} and \ac{OPEX} substantially.

An interesting question for mobile operators is whether the deployed hardware should also be used to supply the vicinity of the train tracks. From a cost perspective, this can be profitable but, simultaneously, it increases the complexity of network planning.

For train operators, it is beneficial to keep control over mobile communications that is when the task for supplying wireless access for their trains is not completely handed off to a mobile operator. For one, the target of mobile operators is mostly to achieve a good coverage with minimal effort. Thus, they will concentrate on lucrative parts of highly frequented routes. The train operator on the other hand has the goal to provide coverage for the whole rail network, with a minimum performance guaranteed everywhere. Thereby, the competitiveness of the train operator is consistently increased, as the additional service is provided comprehensively. Currently, interest among train operators is increasing to employ the hardware for mobile communications also for conveying control data for the trains. Since this aspect will be critical for the security of the system, it will be the operators desire to ensure the \emph{dependability} of the communication system including aspects such as stringent latency constraints. From this perspective, many aforementioned arguments need to be reevaluated. For example, train operators typically do not buy their own licensed frequency bands. On the other hand, freely accessible bands such as the \ac{ISM} bands come with the disadvantage of unpredictable interference, which is not feasible for security relevant systems.
\section{Conclusion}
\label{Sec:Conclusion}

A variety of different aspects in \ac{HST} scenarios was discussed in this paper, with the main focus on the comparison of direct-link and relay approaches.
It was observed that the direct link approach suffers from high penetration loss and a high number of simultaneous handovers which can be overcome by utilizing materials with lower penetration losses (for windows and/or carriage walls), and by using sophisticated mobility management schemes like the moving cell. Moreover, sophisticated scheduler schemes allow to schedule delay-sensitive and best-effort traffic depending on the train's position. If the technical issues with the direct-link approach can be overcome, it grants a less complex system setup and also avoids legal issues with additional hardware installed on the trains. Further considerations of train- and mobile operators increase the complexity of the decision for specific system aspects, with cost issues and other factors coming into play, thus making such scenarios subject to multi-objective optimization.


\section*{Acknowledgements}
This work has been funded by the Christian Doppler Laboratory for Wireless Technologies for Sustainable Mobility, the A1 Telekom Austria AG, and the KATHREIN-Werke KG. The financial support by the Federal Ministry of Economy, Family and Youth and the National Foundation for Research, Technology and Development is gratefully acknowledged.

\bibliographystyle{IEEEtran}\balance
\scriptsize \footnotesize \small
\bibliography{IEEEabrv,References}

\end{document}